\def\year{2021}\relax
\documentclass[letterpaper]{article} % DO NOT CHANGE THIS
\usepackage{aaai21}  % DO NOT CHANGE THIS
\usepackage{times}  % DO NOT CHANGE THIS
\usepackage{helvet} % DO NOT CHANGE THIS
\usepackage{courier}  % DO NOT CHANGE THIS
\usepackage[hyphens]{url}  % DO NOT CHANGE THIS
\usepackage{graphicx} % DO NOT CHANGE THIS
\usepackage{natbib}  % DO NOT CHANGE THIS AND DO NOT ADD ANY OPTIONS TO IT
\usepackage{caption} % DO NOT CHANGE THIS AND DO NOT ADD ANY OPTIONS TO IT
\usepackage{cite}
\usepackage{amsmath,amssymb,amsfonts}
\usepackage{algorithmic}
\usepackage{graphicx}
\usepackage{subfigure}
\usepackage{textcomp}
\usepackage{xcolor}
\usepackage{multirow}
\usepackage{booktabs}

\usepackage[ruled]{algorithm2e}

\title{Cold-start Sequential Recommendation via Meta Learner}
\author {
        Yujia Zheng,\textsuperscript{\rm 1}
        Siyi Liu, \textsuperscript{\rm 1}
        Zekun Li, \textsuperscript{\rm 2, 3}
        Shu Wu \textsuperscript{\rm 4, 5}\\
}
\affiliations {
    \textsuperscript{\rm 1} University of Electronic Science and Technology of China \\
    \textsuperscript{\rm 2} School of Cyber Security, University of Chinese Academy of Sciences \\
    \textsuperscript{\rm 3} Institute of Information Engineering, Chinese Academy of Sciences \\
    \textsuperscript{\rm 4} School of Artificial Intelligence, University of Chinese Academy of Sciences \\    
    \textsuperscript{\rm 5} Institute of Automation and Artificial Intelligence Research, Chinese Academy of Sciences \\
    \{yjzheng19, ssui.liu1022, lizekunlee\}@gmail.com, shu.wu@nlpr.ia.ac.cn \\
}
\begin{document}

\maketitle

\begin{abstract}
This paper explores meta-learning in sequential recommendation to alleviate the item cold-start problem. Sequential recommendation aims to capture user's dynamic preferences based on historical behavior sequences and acts as a key component of most online recommendation scenarios. However, most previous methods have trouble recommending cold-start items, which are prevalent in those scenarios. As there is generally no side information in the setting of sequential recommendation task, previous cold-start methods could not be applied when only user-item interactions are available. Thus, we propose a Meta-learning-based Cold-Start Sequential Recommendation Framework, namely \emph{Mecos}, to mitigate the item cold-start problem in sequential recommendation. This task is non-trivial as it targets at an important problem in a novel and challenging context. Mecos effectively extracts user preference from limited interactions and learns to match the target cold-start item with the potential user. Besides, our framework can be painlessly integrated with neural network-based models. Extensive experiments conducted on three real-world datasets verify the superiority of Mecos, with the average improvement up to 99\%, 91\%, and 70\% in HR@10 over state-of-the-art baseline methods.
\end{abstract}
\section{Introduction}
% Session-based Recommendation System (SRS) has attracted much attention for its highly practical value, especially in some real-world scenarios that concentrated with multitudes of anonymous interactive data. Different from most of the other recommendation tasks that need explicit user demographic profiles, SRS only relies on anonymous user action logs (e.g, clicks) in an ongoing session to predict the user's next action. 

Increasing research interests have been attracted in sequential recommendation due to its highly practical value in online services (e.g., e-commerce), where users' current interest are intrinsically dynamic as the evolving of their historical actions. Accurately recommending the user's next action based only on the sequential dynamic of historical interactions lies in the heart of sequential recommendation. 

Markov Chains (MC) \cite{zimdars01} is a representation of traditional methods, which predicts the user’s next action based on the previous one. Recently, neural network-based methods have become popular due to their strong abilities to model sequential data, such as the methods based on recurrent neural network (RNN) \cite{hidasi2015session, tan2016improved, li2017neural, cui2018mv}, Attention \cite{liu2018stamp, kang2018self}, convolutional neural network (CNN) \cite{tang2018personalized}, and graph neural network (GNN) \cite{wu2019srgnn, zheng2020dgtn}. For example, NARM \cite{li2017neural} employs RNNs with an attention mechanism to capture the main purpose and sequential behavior, then combines them as a representation for the recommendation. SR-GNN \cite{wu2019srgnn} models the session data in the graph structure and utilizes a gated GNN to model complex transitions among item nodes.

% \begin{figure}[ht]
% \centering
% \subfigure[Steam]{\includegraphics[width=0.48\linewidth]{sparsity-with-power/Steam.png}}
% \subfigure[Electronic]{\includegraphics[width=0.48\linewidth]{sparsity-with-power/Electronic.png}}
% % \subfigure[Tmall]{\includegraphics[width=0.33\linewidth]{sparsity-with-power/Tmall.png}}\hspace{-3mm}
% %\vspace{-4mm}
% \caption{Distributions of item frequencies in Steam and Electronic datasets, where green line represents the power law. There is a large portion of items that only have a few interactions.}
% \label{fig: distribution}
% %\vspace{-2mm}
% \end{figure}

These sequential recommendation models mainly rely on user-item interactions to portray user preference and item property.
Therefore, the number of historical interactions largely determines the model performance.
However, new items arrive frequently in those online services, and timely recommending these items to the target users is of high practical value.
As they rarely have enough known interactions with users, existing models have trouble in recommendation towards those items.
Therefore, cold-start problem needs to be solved in sequential recommendation scenarios.
% In need of timely recommending these items to the target users, cold-start problem is of high practical value in those scenarios. Existing sequential recommendation models show strong capabilities in capturing user-item interaction signals, but they have difficulties in recommendation toward cold-start items. Because the basic paradigm of previous work is to portray the user's preference as a vector of its historical interactions, previous sequential recommendation methods rely heavily on the number of known interactions between items and users. Consequently, the number of interactions and historical sequences largely determines the upper-bound of the model, which leads to poor performance on cold-start recommendation

The basic idea behind existing cold-start methods in the general recommendation is to incorporate side information, such as user and item attributes (e.g., user profile, item ratings or descriptions) \cite{barjasteh2016cold, saveski2014item} and knowledge from other domains \cite{hu2018conet}. Recently, some meta-learning approaches have been proposed to tackle it \cite{vartak2017meta, yao2019learning, lee2019melu, du2019sequential, dong2020mamo}, but they still cannot get rid of these side information. For example, $s^{2}$Meta \cite{du2019sequential} relies on cross-domain knowledge transferring. Vartak et al. \cite{vartak2017meta} utilize item ratings to make the twitter recommendation. MAMO \cite{dong2020mamo} predicts the rating for new items with the help of user profile and item descriptions. However, because there is generally no side information in the setting of sequential recommendation task, previous cold-start methods, including those based on meta-learning, cannot be applied in these scenarios. Therefore, the cold-start problem in sequential recommendation remains unexplored. 

To alleviate the item cold-start problem in sequential scenarios given only sparse interactions, we propose a MEta-learning-based COld-start Sequential recommendation framework (\emph{Mecos}). Our framework can be painlessly integrated with neural network-based sequential recommendation models. We first design a sequence pair encoder to effectively extract user preference. Then a matching network is applied to match the query sequence pair to the support set corresponding to the candidate cold-start item. Moreover, we employ the meta-learning based gradient descent approach for parameter optimization. Once trained, the learned metric model can be applied to make recommendations for new items without further fine-tuning. Extensive experiments show the superiority of Mecos in dealing with cold-start items in sequential recommendation. 
 
The main contribution of this work is threefold. First, we propose Mecos framework for sequential scenarios to alleviate the item cold-start problem. To the best of our knowledge, this is the first work to tackle this problem. The task is non-trivial as it aims at a classical problem in a novel and challenging context, where the only available data is the user-item interaction. Second, our framework can be easily integrated with existing neural network-based sequential recommendation models. And once trained, Mecos can be adapted to new items without further fine-tuning. Lastly, we compare Mecos with state-of-the-art methods on three public datasets. Results demonstrate the effectiveness of our framework. And a comprehensive ablation study is conducted to analyze the contributions of the key components.

\section{Related Work}
\subsection{Sequential Recommendation}

% \textbf{Conventional methods. } 
% Matrix Factorization (MF) \cite{koren2009matrix} is widely used in recommendation system. MF factorizes a user-item rating matrix to get latent user vectors to represent general interests and make recommendations. Another widely used approach is the item-based neighborhood method, which only considers the last item in a given sequence and recommends the most similar items in terms of their co-occurrence in other sequences.  Different from the methods mentioned above, the sequential methods based on Markov Chains (MC) focuses on the data sequentiality. Shani et al. \cite{shani2005mdp} apply Markov Decision Processes (MDPs) to extract sequential patterns from user sequential action logs and make predictions. Moreover, Rendle et al. \cite{rendle2010factorizing} propose FPMC method, which combines MC and traditional MF in a three-dimensional tensor factorization approach to make next-basket recommendations.

% \textbf{Deep-learning-based methods. }
Neural networks-based recommenders \cite{wu2020tfnet, li2020dynamic} have attracted much attention due to their effect in modeling sequential data. RNN with a gated recurrent unit \cite{hidasi2015session} achieves significant improvement compared with conventional methods. To further improve it, an encoder-decoder model (NARM) applying attention mechanism \cite{li2017neural} is proposed to extract more representative features. Besides these RNN-based methods, CNN is used in Caser \cite{tang2018personalized} to learn local features of a sequence "image" using convolutional filters. Moreover, to explicitly model the impact of items of different time steps on the current decision, attention-based models are also developed in sequential recommendation, such as STAMP and SASRec \cite{liu2018stamp, kang2018self, liu2020long}.
% Liu et al. \cite{liu2018stamp} applied the short-term attention priority model (STAMP) to model current and general user interests. Kang et al. \cite{kang2018self} leveraged self-attention to adaptively capture long-term semantics based on relatively few actions. Liu et al. \cite{liu2020long} proposed a preference mechanism to improve the coverage of long-tail items. 
As GNN becomes popular, it is also employed to model sequential data into graph-structured data to capture complex items transitions \cite{wu2019srgnn, yu2020tagnn}. Recent works \cite{zheng2019balancing, zheng2020dgtn} introduced cross-session item transitions. Different from these methods, we explore few-shot learning to alleviate the item cold-start problem in sequential recommendation.

\subsection{Cold-start Recommendation}
The common solution to address cold-start recommendation is to utilize side information, such as auxiliary and contextual information \cite{barjasteh2016cold, saveski2014item} and knowledge from other domains \cite{hu2018conet}. Traditional content-based methods augment the data with user or item features. Saveski and Mantrach \cite{saveski2014item} proposed a local collective embedding learning method to exploits items' properties and user preferences. DecRec \cite{barjasteh2016cold} decouples the rating sub-matrix completion and knowledge transduction from ratings to exploit the side information. Besides, based on the assumption that the same user or item can be located in different scenarios, transferring-based methods \cite{hu2018conet, kang2019semi} use cross-domain knowledge to mitigate cold-start problems in a new scenario.

% There are two popular types of approaches for meta-learning. The first one is the gradient-based approach, which aims to quickly optimize the model parameter given the gradient of few-shot examples \cite{finn2017model,finn2018probabilistic,lee2018gradient, yao2019hierarchically} or learn a better initialization of parameter that could be well adapted for future tasks \cite{ravi2017optimization}.
% % One example is the MAML \cite{finn2017model} which explicitly train the parameters that increase the generalization ability for new tasks. 
% Another type is the metric-based approach, which learns a generalized metric and corresponding matching function from training tasks \cite{snell2017prototypical,vinyals2016matching,bertinetto2019meta,sung2018learning}.
% For example, matching networks \cite{vinyals2016matching} learns a network that maps the input example with small labeled support set to obviate the expensive fine-tuning for new types. 

% Previous meta-learning study mainly focus on vision \cite{sung2018learning}, imitation learning \cite{duan2017one}, spatial-temporal prediction \cite{yao2019learning} and continuous control \cite{yoon2018bayesian}. 
\subsubsection{Meta-learning in Cold-start Recommendation}
Recently, meta-learning for recommender systems has been attracting attention. Most of these works focus on the recommendation scenarios with few training samples because it is natural to turn these tasks into few-shot learning problems. 
For example, Vartak et al. \cite{vartak2017meta} propose two network architectures based on a meta-learning strategy to predict the user's rating for twitters based on historical item ratings. $s^{2}$Meta \cite{du2019sequential} initializes the recommender with a sequential learning process based on cross-domain knowledge transferring. Li et al. \cite{li2019zero} formulate cold-start recommendation as a zero-shot learning task with user profiles. MeLU \cite{lee2019melu} integrates both user and item attributes and identifies reliable evidence candidates for customized preference estimation. Also relying on user and item side information, MAMO \cite{dong2020mamo} uses two designed memory matrices to avoid the local optima for users with a specific pattern. However, as they all rely on side information (e.g., user profile, item attributes, and cross-domain knowledge) to leverage limited training data, they cannot be applied in the sequential recommendation, where the only data that is always available is the user-item interaction. Different from them, our work introduces a general framework to alleviate cold-start problem based only on user-item interactions.

\section{Preliminary}

\subsubsection{Problem Formulation}
In sequential recommendation, a dataset can be represented as a collection of user-item sequences. Let $\mathcal{U}$ and $\mathcal{V}$ represent the user set and item set respectively. And $\zeta_i = (v_{i,1}, v_{i,2}, ..., v_{i,n})$ denotes the interaction sequence generated by user $u_i \in \mathcal{U}$ in the chronological order. The aim of sequential recommendation model is to recommend the next item $v_{i,n+1}$ that the user $u_i$ will interact with based on $\zeta_i$. Thus, it is equal to predict next item $v_{i,n+1}$ based on the query sequence pair $(\zeta_i, \text{?}\footnote{
The question mark ``\text{?}" represents an unknown next-click item that needs to be recommended.})$. To achieve that, the model will generate a score for each candidate item in the item set $\mathcal{V}$. Then items with a top-$p$ recommendation score will be recommended as the model’s output. 

% \text{?}\footnote{
% The question mark ``\text{?}" represents an unknown next-click item that needs to be recommended.}

In our task, we have a set of cold-start items $\mathcal{Y} \subset \mathcal{V}$, that every single item in $\mathcal{Y}$ only has a few interactions with users in the dataset. Different from other sequential recommendation models that rely on rich training instances, our goal is to recommend cold-start items with few-shot examples. We use $\mathcal{Y}_\text{train}$ and $\mathcal{Y}_\text{test}$ to represent the training and testing instances, where $\mathcal{Y}_\text{test} \cap \mathcal{Y}_\text{train} = \phi$.

% anonymous triples $(s, l, n) \subseteq \mathcal{S} \times \mathcal{V}$, where $\mathcal{S}$ and $\mathcal{V}$ are the session set and item set. And $s$ in $\mathcal{S}$ contains a sequence of items $[v_1, v_2, ..., v_{n-1}]$ and the last item $ l = v_n$, where $v_n \in \mathcal{V}$. $n$ represents the next click item that will be treated as the ground truth label of the session 
% $s$. The goal of SRS is to recommend next item $n = v_{n+1}$ based on the query session $(s, l, ?)$. Unlike the previous session-based recommendation work that assumes all triples in the dataset are available for training, we set a challenging and representative scenario that only one training triple is available for each last item $l$ that represents the user's current preference. Given a query session $(s, l, ?)$ and an example triple $(s_0, l, n_0)$, the goal is to generate a score for each candidate item in the item set $\mathcal{V}$. Then items with a top-$N$ recommendation score will be recommended as our model’s output. 

\begin{figure}[htp]
\centering
\includegraphics[width=\linewidth]{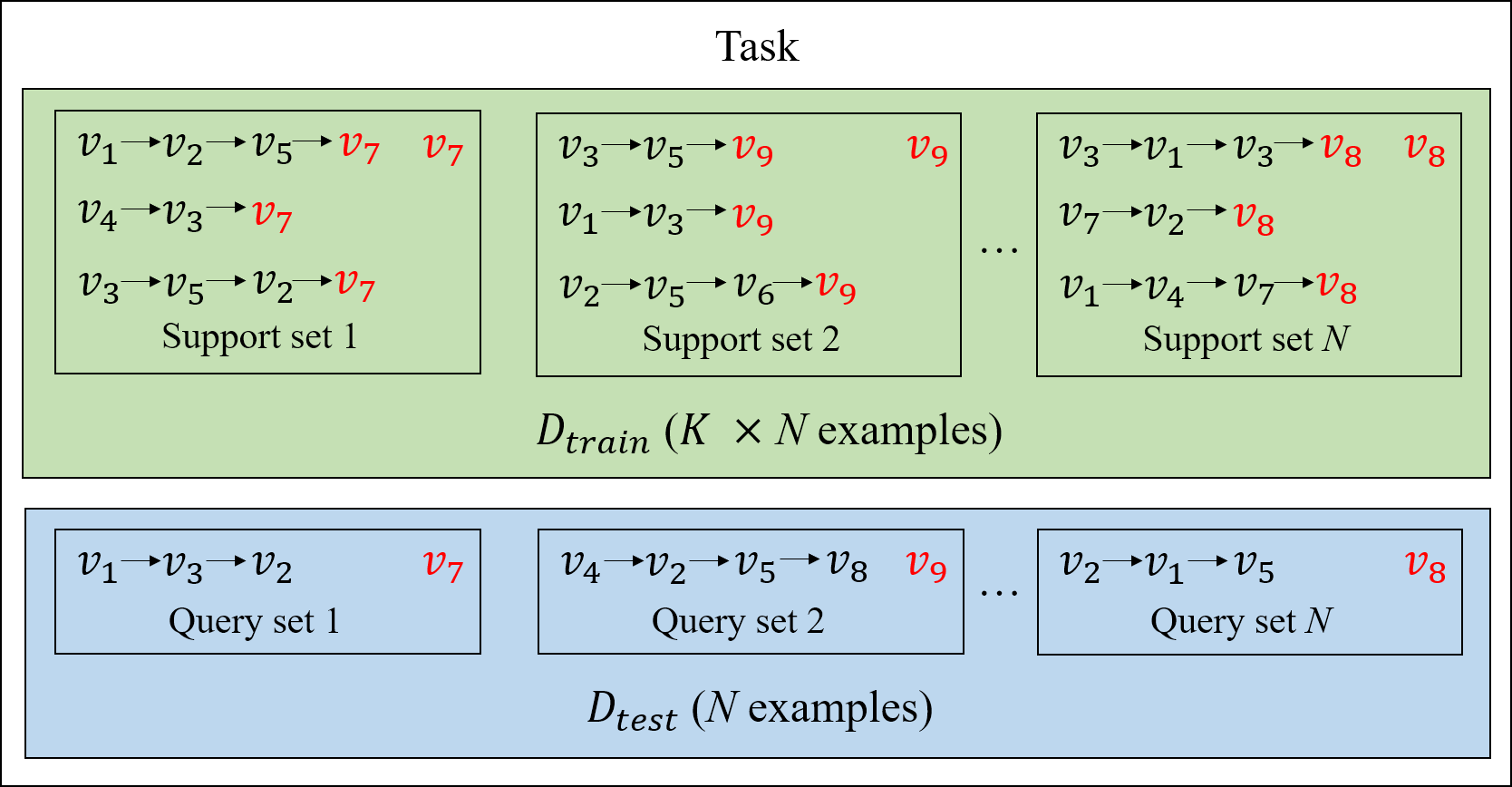} % Reduce the figure size so that it is slightly narrower than the column. Don't use precise values for figure width. This setup will avoid overfull boxes. 
\caption{Single task with $K = 3$ and $N$ different support sets. Each query set is the query sequence pair that needs to be matched with a candidate cold-start item (red) corresponding to the support set. When $K = 3$, we have three training instances for our target item. And $N$ is the size of candidate cold-start items set.}
\label{fig: task_example}
\vspace{-2mm}
\end{figure}

\subsubsection{Meta-training}
% In Few-Shot Learning, we have a set of users interaction sequences that their next interacted items belong to sparse items, which we denoted as $\mathcal{Y}$ and treated as our all classes in our experiment. 

Following the standard meta-learning settings, we train our model based on a set of training tasks $\mathcal{T}_\text{meta-train}$. To create a training task $T_i \in \mathcal{T}_\text{meta-train}$ (Figure \ref{fig: task_example}), we first sample $N$ next-click items from $\mathcal{Y}_\text{train}$. For each of those $N$ items, we sample $K$ sequence pairs, where each pair $(\zeta_i, v_{i,n+1})$ represents a sequence $\zeta_i$ and its ground-truth next-click item $v_{i,n+1}$, as the support set. As the cold-start item is recommended to one user at each time, the query set is represented by a single query sequence pair $(\zeta_i, \text{?})$. We denote those $N$ support/query sets in task $T_i$ as $D_i^\text{train}$ and $D_i^\text{test}$, respectively. Then the query set will match over all support sets to calculate the similarities between them. The similarity score will be handled as the recommendation score of the candidate next-click item in the corresponding support set. Since we employed the data augmentation on the datasets as previous methods \cite{liu2018stamp, wu2019srgnn} (e.g., a sequence $(v_0, v_1, v_2, v_3)$ is divided into three successive sequences:  $(v_0, v_1)$, $(v_0, v_1, v_2)$, $(v_0, v_1, v_2, v_3)$), the candidate next-click items are actually distributed in all positions of the original sequence thus they are equal to the candidate items. Then the loss can be calculated by the ground-truth next-click item and the recommended one. We will elaborate these in the next sections.

\subsubsection{Meta-testing}
After training, the model can make predictions for the task with $N$ new ground-truth items from $\mathcal{Y}_\text{test}$, which is the $\emph{meta-testing}$ step. These meta-testing ground-truth items are unseen from the meta-training. The same as meta-training, each meta-testing task also has its few-shot training data and testing data. And these tasks form the meta-test set $\mathcal{T}_\text{meta-test}$. Moreover, we randomly leave out a subset of labels in  $\mathcal{Y}_\text{train}$ to generate the validation set $\mathcal{T}_\text{meta-valid}$. 

Now we can define our total meta-learning task set as $\mathcal{T} = \mathcal{T}_\text{meta-train} \bigcup \mathcal{T}_\text{meta-test} \bigcup \mathcal{T}_\text{meta-valid}$.
% In addition, we create a subset of the dataset, with all tasks in $\mathcal{T}$ removed, for the training of pre-trained item embeddings.

\subsubsection{Integrating with Existing Methods}

% For those sequences that their ground-truth labels belong to short-head, denoted as $\mathbf{S}^h$, we treat them as pre-train data to obtain a more informative embedding for those high-frequent items. To make sure the $mathcal{Y}$ will be cold-started, sequences that contain items belong to $\mathcal{Y}$ will be discarded. 

With the input of pre-trained item embeddings generated by existing sequential recommendation models, Mecos can be easily integrated with these methods to improve their performance in cold-start scenarios. Based on our setting, tasks that their ground-truth next-click items have rich interactions will be excluded from $\mathcal{T}$. And these data will be selected for the training of pre-trained item embeddings. To ensure those ground-truth items will not be seen before meta-learning tasks, sequences in pre-trained datasets will be removed when they contain the ground-truth item in $\mathcal{T}$.

It is noteworthy that we are not introducing more complexity or training with more data based on the baseline models. The purpose of pre-training is to make sure that both models with or without Mecos have a similar quality of embeddings for items with rich interactions. And all models are training with the same amount of data. Under these circumstances, we can make a fair comparison of recommendation performance toward cold-start items. 

% After training, we input pre-trained item embeddings to our proposed framework Mecos, thus We utilize state-of-the-art methods in sequential recommendation, such as Caser, SASRec and SR-GNN, as backbone models to obtain pre-trained item embeddings, which are further used as the input fof our proposed Mecos.

% \begin{table}
% \centering
% \scriptsize
% \caption{Notations.}\label{tab:notation}
% \vspace{-3mm}
% \resizebox{0.45\textwidth}{!}{
% \begin{tabular}{ll}
%     \toprule
%     \textbf{Notation} & \textbf{Representation} \\
%     \midrule
%     $\mathbf{U}$ & user set \\
%     $\mathbf{V}$ & item set \\
%     $\mathcal{Y}$ & cold start items set \\
%     $\mathcal{Y}_\text{train}$ & cold start items set of meta-training \\
%     $\mathcal{Y}_\text{test}$ & cold start items set of meta-testing \\
%     $T_i$ & a single task \\
%     $D_i^\text{train}$ & training data in $T_i$ \\
%     $D_i^\text{test}$ & testing data in $T_i$ \\
%     $\mathcal{T}$ & total task set \\
%     $\mathbf{h}_\mathrm{s}$ & support set representation \\
%     $\mathbf{h}_\mathrm{q}$ & query set representation  \\
%     $\hat{\mathbf{y}}_i$ & the recommendation score of query set $i$ \\
%     \bottomrule
% \end{tabular}}
% \vspace{-5mm}
% \end{table}

\section{Mecos}
\begin{figure*}[htp]
\centering
\includegraphics[width=2\columnwidth]{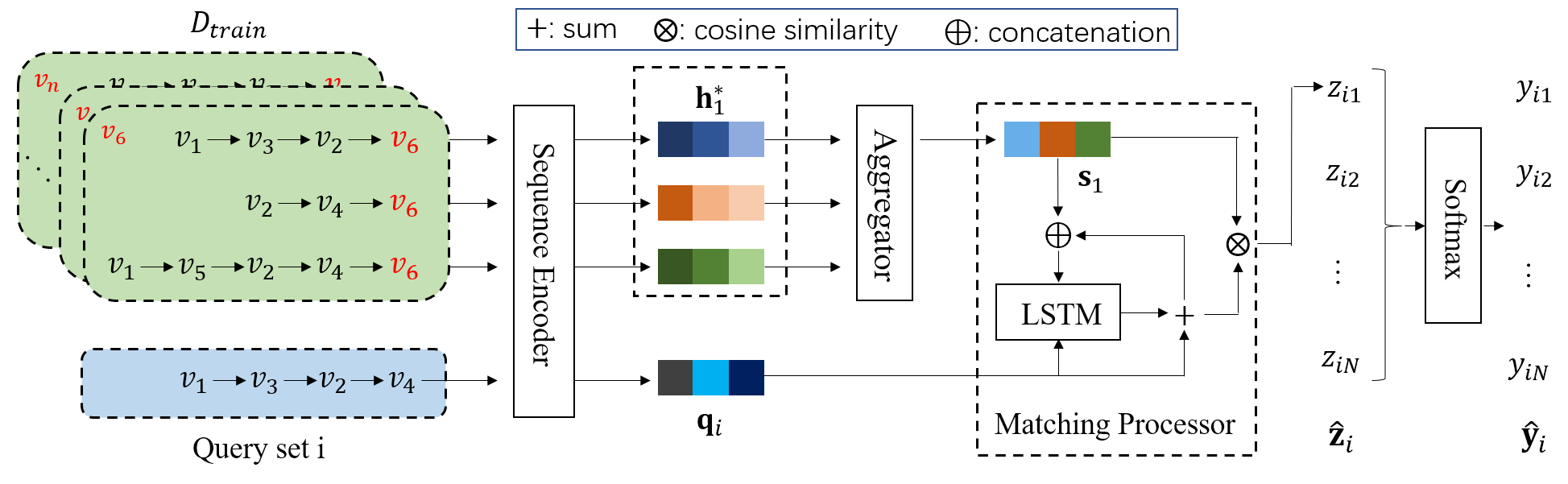} % Reduce the figure size so that it is slightly narrower than the column. Don't use precise values for figure width. This setup will avoid overfull boxes. 
\caption{The framework of Mecos. The green square represents a support set that contains few-shot training sequences of a candidate cold-start item (red). Our goal is to recommend cold-start items to the query sequence (blue square). It first generates sequence pair representations (Eq. \ref{eq1}, \ref{eq2}), then aggregates few-shot sequence representations to generate support set representation (Eq. \ref{eq3}). Finally, it employs a matching network (Eq. \ref{eq4}) to compute similarity scores between the query set and all support sets (Eq. \ref{eq5}). After matching the $i$-th query set with $N$ different support sets, we can get a vector of similarity scores $\hat{\mathbf{z}}_i$. Then a softmax function is applied to generate the $
i$-th query set's final recommendation score of each candidate item, denoted as $\hat{\mathbf{y}}_i$.}
\label{fig:framework}
\vspace{-2mm}
\end{figure*}

In this section, we elaborate on the details of Mecos (Figure \ref{fig:framework}). With the pre-trained item embeddings as inputs, Mecos first encodes sequence pairs into representation vectors, then aggregates them to generate the support/query set representations. After that, a matching network is employed to match the query set with the support set, which is corresponding to a candidate cold-start item, as the recommendation result.

% We here introduce two key steps of our proposed framework: 1. \emph{Encoding Support and Query Set}, 2. \emph{Matching Support and Query Set}. The objective function and training process of our meta-learning tasks will also be clarified. Figure \ref{fig:framework} shows the framework of Mecos. 
% With the pre-trained item embeddings as inputs, Mecos first uses them to generate the embeddings of each sequence pairs, then aggregates them to generate the support/query set representations. Finally, a matching network is employed to match query set with the support set, which is corresponding to a candidate cold-start item, as the recommendation.

\subsubsection{Encoding Support and Query Set}

Due to the inconsistency of sequence length among different sequences, we need to obtain a uniform sequence pair representation for matching. So the Sequence Pair Encoder is designed to get the representation vector $\textbf{s}_{i}$ of the support sequence pair $(\zeta_i, v_{i,n+1})$ and $\textbf{q}_{i}$ of the query sequence pair $(\zeta_i, \text{?})$, where $\zeta_i$ contains a sequence of items interacted by user $u_i$. $\mathbf{v}_{i,j} \in \mathbb{R}^{d}$ denotes the pre-trained embedding of $j$-th item in 
$\zeta_i$, and $\mathbf{v}_{i,n+1} \in \mathbb{R}^{d}$ is the next-click item embedding. 

The first step is obtaining the sequence representation $\mathbf{s}$. Since the last-click item in the sequence basically represents user's current interest, we incorporate it into the sequence embedding $\mathbf{s}$ through self-attention:
\begin{equation} \label{eq1}
    \begin{split}
    \begin{aligned}
    \mathbf{e}_{j} &= \mathbf{p}^{T} \left(\mathbf{W}^{1}_a \mathbf{v}_{i,n}+\mathbf{W}^{2}_a \mathbf{v}_{i,j}+\mathbf{W}^{3}_a \mathbf{v}_{i,avg}+\mathbf{b} \right), \\
    \alpha_{j} &= \frac{exp(\mathbf{e}_{j})}{\sum\nolimits_{k=1}^{n} exp(\mathbf{e}_{k})}, \\
    \mathbf{R}({\zeta}_{i}) &= \sum\nolimits_{j=1}^{n} \alpha_{j} \mathbf{v}_{i,j},
    \end{aligned}
    \end{split}
\end{equation}
where $\mathbf{R}({\zeta}_{i})$ is the representation of sequence $\zeta_i$, $\mathbf{p}^{T} \in \mathbb{R}^{d}$ is a projection vector, $\mathbf{W}^{1}_a, \mathbf{W}^{2}_a, \mathbf{W}^{3}_a \in \mathbb{R}^{d \times d}$ are learnable weighted parameters, $\mathbf{v}_{i,avg} = \sum\nolimits_{j=1}^{n} \left(\mathbf{v}_{i,j}\right) / n$ is the average item embedding, and $\mathbf{b}$ is a bias vector. 

Then, for sequence pairs in the support set, we combine the next-click item embedding with the sequence representation vector $\mathbf{\zeta}_i$ to generate sequence pair representation $\mathbf{h}$. We utilize a feed-forward network with standard residual connection to better merge the representations:
\begin{equation} \label{eq2}
    \begin{split}
    \mathbf{h}_{i}^{0} &= \left[\mathbf{R}({\zeta}_{i}); \mathbf{v}_{i,n+1}\right], \\
    \mathbf{h}_{i}^{l} &= \operatorname{ReLU}\left(\mathbf{W}^{l}\mathbf{h}_{i}^{l-1} + \mathbf{b}^{l} \right), \\
    \mathbf{h}_{i} &= \mathbf{h}_{i}^{0} + \mathbf{h}_{i}^{l},
    \end{split}
\end{equation}
where $\mathbf{h}_{i}$ is the sequence pair representation. $\mathbf{W}^{l} \in \mathbb{R}^{2d \times 2d}$ and $\mathbf{b}^{l} \in \mathbb{R}^{2d}$ denote the weight and bias for layer $l$, respectively. $\operatorname{ReLU}(z)$ is a non-linear activation function.

% In addition, if $K > 1$, we use the mean pooling to aggregate the K support set sequence tuple representation $\mathbf{h^1}, \mathbf{h^2}, ... ,\mathbf{h^K}$ to one single representation $\mathbf{h_s}$ for each ground-truth item. 

After encoding $K$ sequence pairs in the support set into $\mathbf{h}^{*}_{i}=\{\mathbf{h}_{i, 1}, \mathbf{h}_{i, 2}, ... ,\mathbf{h}_{i, K} \}$, we aggregate them into the support set representation $\mathbf{s}_i$:
\begin{equation} \label{eq3}
    \begin{split}
    \mathbf{s}_i &= \textit{Aggr}(\mathbf{h}^{*}_{i}),
    \end{split}
\end{equation}
% \begin{equation} \label{eq3}
%     \begin{split}
%     \mathbf{h}_{\mathrm{s}, i} &= \textit{Aggr}(\mathbf{h}^{*}_{i}),
%     \end{split}
% \end{equation}
where $\textit{Aggr}(\cdot)$ could be mean pooling, max pooling, feed-forward neural network, etc. To keep the framework simple, we choose the mean pooling layer as $\textit{Aggr}(\cdot)$.

Because the cold-start item is recommended to one query at each time, the representation of the query set $\mathbf{q}_i$ is equal to that of the query sequence pair $(\zeta_i, \text{?})$. Moreover, for the sequence pair in the query set, the next-click item should be inaccessible to prevent data leakage. Thus, in generating the query pair embedding, $\mathbf{h}_{i}^{0}$ in Equation \ref{eq2} should be rewritten as $\mathbf{h}_{i}^{0} = \mathbf{W}_{q}\mathbf{R}({\zeta}_{i})$, where $\mathbf{W}_{q} \in \mathbb{R}^{2d \times d}$ denotes a projection matrix. And the output of Equation \ref{eq2} will be the query set representation $\mathbf{q}_i$.

% After thateq, we concatenate representations of all sequence pairs into a support set representation $\mathbf{H_s} \in \mathbb{R}^{N \times 2d}$

% Moreover, for sequence pairs in the query set, the next-click item should be inaccessible to prevent data leakage. So the $\mathbf{h}_0$ in Equation \ref{eq2} will be rewritten as $\mathbf{h_0} = \mathbf{W}_{q}\mathbf{s}$, where $\mathbf{W}_{q} \in \mathbb{R}^{2d \times d}$ denotes a projection matrix. And with the same other processes in generating support set representation $h_s$, we can obtain the query set representation $h_q$.
 
% \begin{equation} \label{eq5}
%     \begin{split}
%     \mathbf{h_0} &= \left[\mathbf{s}; \mathbf{s}\right],
%     \end{split}
% \end{equation}
% The final query pair representation is denoted as $\mathbf{h_q}$ and the representation of the query set is denoted as $\mathbf{H_q} \in \mathbb{R}^{NL \times 2d}$.

\subsubsection{Matching Support and Query Set}

% After getting support pair embedding set $\mathbf{H_s}$ and query pair embedding set $\mathbf{H_q}$, for each $\mathbf{h_q}$, the next step is to evaluate the similarities between $\mathbf{h_q}$ and . 

After encoding support and query sets, we get the support set representation set $\mathbf{S}$ and query set representation set $\mathbf{Q}$. Then we need to match the $i$-th query set representation $\mathbf{q}_{i} \in \mathbf{Q}$ with the $j$-th support set $\mathbf{s}_{j} \in \mathbf{S}$. There are lots of popular similarity metrics, such as Euclidean distance and Cosine similarity. These meatrics simply measures the distance between two vectors in the embedding space, which does not represent relativity in real life. Thus we need to learn a general distance function to capture the real connection in the application scenarios. To address this, we leverage a recurrent matching processor \cite{vinyals2016matching} for multi steps matching. Once trained, our model can be used to any new item types without re-training. The $t$-th step could be defined as following:
\begin{equation} \label{eq4}
\begin{split}
\hat{\mathbf{q}}_{i}^{t}, \mathbf{c}^{t} &= \mathrm{LSTM}\left(\mathbf{q}_{i},\left[\mathbf{q}_{i}^{t-1}; \mathbf{s}_{j}\right], \mathbf{c}^{t-1}\right), \\
\mathbf{q}_{i}^{t} &= \hat{\mathbf{q}}_{i}^{t} + \mathbf{q}_{i},
\end{split}
\end{equation}
where LSTM is a LSTM cell \cite{hochreiter1997long} with input $\mathbf{q}_{i}$, hidden state $\left[\mathbf{q}_{i}^{t-1} ; \mathbf{s}_{j}\right]$ and cell state $\mathbf{c}^{t-1}$. The last hidden state $\mathbf{q}_{i}^{t}$ after $t$ steps is the refined embedding of the query pair.

Then the similarity score is measured by cosine similarity:
\begin{equation} \label{eq5}
\begin{array}{c}

z_{ij}=\displaystyle\frac{\mathbf{q}_{i}^{t} \mathbf{s}_{j}}{\left\|\mathbf{q}_{i}^{t}\right\| \times\left\|\mathbf{s}_{j}\right\|},

\end{array}
\end{equation}
where $z_{ij}$ denotes the similarity between query set representation $\mathbf{q}_{i}$ and support set representation $\mathbf{s}_{j}$. According to the setting, every support set corresponds to a candidate item. Thus, $z_{ij}$ represents the recommendation score of the candidate item tied with the $j$-th support set and $\hat{\mathbf{z}}_i \in \mathbb{R}^{N}$ denotes the set of all candidate item scores for query pair $(\zeta_i, ?)$.
% For every , and $\hat{\mathbf{z}_i} \in \mathbb{R}^{N}$ to the s $\mathbf{h_{q, i}}$ over $N$ candidate support pairs, which also could represent the recommendation score over $N$ candidate ground-truth items.

\begin{algorithm}[htb] 
\caption{Meta-training process} 
\label{alg:Framwork} 
\begin{algorithmic}[1] %这个1 表示每一行都显示数字
\REQUIRE  %算法的输入参数：Input
% \begin{array}{l}{\text{Meta-training task set $\mathcal{T}_{train}$};} \\ 
% {\text {Pretrained item embeddings} $\mathbf{V}_p$ ;} \\ 
% {\text {Initial model parameters } $\Theta$;} 
% \end{array}
% \end{array}
Meta-training task set $\mathcal{T}_\text{meta-train}$, pre-trained item embedding $\mathbf{v}_{i}$, initial model parameters $\Theta$
\ENSURE %算法的输出：Output
\text {Model's optimal parameters } $\Theta^{*}$ 
\WHILE {\emph{unfinished}}
\STATE \text {Shuffle tasks in } $\mathcal{T}_\text{meta-train}$
\label{code:fram:extract}%对此行的标记，方便在文中引用算法的某个步骤
\FOR {$\textbf{T}_i \in \mathcal{T}_\text{meta-train}$}
\label{code:fram:trainbase}
% \STATE \text{Create} $D_i^{train}$ \text{and} $D_i^{test}$. 
\STATE \text{Sample support sets and query sets}
\STATE Generate support and query set representations $\mathbf{S}, \mathbf{Q}$
\FOR {$\mathbf{q}_{i} \in \mathbf{Q}$}
\STATE \text{Match with support set representations in} $\mathbf{S}$
\STATE \text{Accumulate the loss by Eq. \ref{eq7}}
\ENDFOR
\STATE \text{Update parameters $\Theta$ by Adam optimizer}
\label{code:fram:classify}
\ENDFOR
\ENDWHILE
\RETURN $\Theta^{*}$
\end{algorithmic}
\end{algorithm}

\subsubsection{Objective Function and Model Training}

After obtaining $\hat{\mathbf{z}}_i$, we apply a softmax function to generate the output vector $\hat{\mathbf{y}}_i$ of the model:
\begin{equation} \label{eq15}
    \begin{split}
    \hat{\mathbf{y}}_i=\operatorname{softmax}(\hat{\mathbf{z}}_i),
    \end{split}
\end{equation}
where $\hat{\mathbf{y}}_i \in \mathbb{R}^{N}$ denotes the probabilities of items becoming the next-click item in the query sequence $\zeta_i$.

The training process is shown in Algorithm \ref{alg:Framwork}. For each task, we apply cross-entropy as the loss function:
\begin{equation} \label{eq7}
    \begin{split}
    \mathcal{L}_{T} = -\sum_{i=1}^{N}\sum_{j=1}^{N} y_{ij} \log \left(\hat{y}_{ij}\right)+\left(1-y_{ij}\right) \log \left(1-\hat{y}_{ij}\right),
    \end{split}
\end{equation}
% \begin{equation} \label{eq16}
%     \begin{split}
%     \mathcal{L}(\hat{\mathbf{y}})=-\sum_{i=1}^{m} \mathbf{y}_{i} \log \left(\hat{\mathbf{y}}_{i}\right)+\left(1-\mathbf{y}_{i}\right) \log \left(1-\hat{\mathbf{y}}_{i}\right),
%     \end{split}
% \end{equation}
where $\mathbf{y}_i$ denotes the one-hot vector of the ground truth item for $i$-th query pair. And $y_{ij}$, $\hat{y}_{ij}$ represent the $j$-th element in vector $\mathbf{y}_i$ and $\hat{\mathbf{y}}_i$, respectively. Finally, our model is trained by Back-Propagation Through Time (BPTT) algorithm.

\section{Experiments}
We conduct experiments on three real-world datasets to study our proposed framework. We aim to answer the following research questions: 

\textbf{RQ1}: How does integrating state-of-the-art models into Mecos perform as compared with the original models? 

\textbf{RQ2}: How do different components in Mecos affect the framework?

\textbf{RQ3}: What are the influences of different settings of hyper-parameters (matching steps and support set size)? 

\textbf{RQ4}: How do the representations benefit from Mecos?

% \begin{itemize}
%     \item {\textbf{RQ1}: How does integrating state-of-the-art models into Mecos perform as compared with the original models?
%     }
%     \item {\textbf{RQ2}: How do different components in Mecos affect the framework?}
%     \item {\textbf{RQ3}: What are the influences of different settings of hyper-parameters (matching steps, support set size and hidden dimensionality) for Mecos?}
%     \item {\textbf{RQ4}: How do the representations benefit from the Mecos?}
% \end{itemize}
\subsection{Experimental Setup}
\textbf{Datasets. } We use three public benchmark datasets for experiments. The first one is based on \textbf{Steam}\footnote{\url{https://cseweb.ucsd.edu/~jmcauley/datasets.html\#steam\_data}}\cite{kang2018self}, which contains user's reviews of online games from Steam, a large online video game platform. The second one is based on Amazon \textbf{Electronic}\footnote{\url{http://jmcauley.ucsd.edu/data/amazon}}, which is crawled from amazon.com by \cite{mcauley2015image}. The last one is based on \textbf{Tmall}\footnote{\url{https://tianchi.aliyun.com/dataset/dataDetail?dataId=47}} from IJCAI-15 competition. This dataset contains user behavior logs in Tmall, the largest e-commerce platform in China. We apply the same preprocessing as \cite{kang2018self,tang2018personalized}. 
% For all datasets, we use the timestamp to order the sequence of interactions. And all reviews or ratings are considered as an interaction no matter their content. 

\begin{table}
\centering
\scriptsize
\caption{Statistics of three datasets.}\label{tab:datastat}
\vspace{-3mm}
\resizebox{0.45\textwidth}{!}{
\begin{tabular}{lrrr}
    \toprule
    \textbf{Datasets} & \textbf{Steam} & \textbf{Electronic} & \textbf{Tmall} \\
    \midrule
    \# sequences & $358,228$ & $443,589$ & $918,358$ \\
    \# items & $11,969$ & $63,002$ & $624,221$ \\
    \# ground-truth items & $9,496$ & $50,331$ & $219,560$ \\
    Propotion of meta-sequences  & $0.12$ & $0.31$ & $0.20$ \\
    \bottomrule
\end{tabular}}
\vspace{-5mm}
\end{table}

% \begin{itemize}
% \item {\textbf{Steam}\footnote{https://cseweb.ucsd.edu/~jmcauley/datasets.html\#steam\_data}: This dataset is introduced in \cite{kang2018self}. The dataset contains user's reviews of online games from Steam, a large online video game platform.}
% \item {Amazon \textbf{Electronic}\footnote{http://jmcauley.ucsd.edu/data/amazon/}: A series of datasets cralwed from Amazon.com by \cite{mcauley2015image}. They split the dataset by top-level product categories on Amazon.com. We adopt the "Electronic" category in this work. This dataset is known for its variety and sparsity.}
% \item {\textbf{Tmall}\footnote{https://tianchi.aliyun.com/dataset/dataDetail?dataId=47}: This datasets contains user bahavior logs in Tmall, the largest e-commerce platform in China. And it is introduced in the IJCAI-15 competition.}
% \end{itemize}

Same as \cite{liu2018stamp, tan2016improved}, the data augmentation strategy is employed on all datasets (e.g., a sequence $(v_0, v_1, v_2, v_3)$ is divided into three successive sequences:  $(v_0, v_1)$, $(v_0, v_1, v_2)$, $(v_0, v_1, v_2, v_3)$). And that strategy is proved to be effective by previous studies \cite{liu2018stamp, tan2016improved}. According to Pareto Principle (i.e., 80/20 rule), we select the 20 percent next-click items with the least number of interactions as few-shot tasks. And sequences corresponding to them are denoted as meta-sequences. Besides, the proportion of training, validation and testing tasks is $7:1:2$. The statistics of data are listed in Tabel \ref{tab:datastat}.

% The first one, called Yoochoose-One, is build based on Yoochoose\footnote{http://2015.recsyschallenge.com/challenge.html} dataset. The YOOCHOOSE dataset is released by the RecSys challenge 2015, which records click sequences (item views, purchases) for a period of six months. Following previous works, we only choose the most recent 1/4 sessions to carry out our study.

% The second one created from Diginetica\footnote{http://cikm2016.cs.iupui.edu/cikm-cup} dataset is named as Diginetica-One. The Diginetica dataset is published by CIKM Cup 2016, in which we only select the transaction data for experiments.

% For fair comparison, we apply the same preprocessing as \cite{liu2018stamp, wu2019srgnn}. Then we split the sessions into two groups based on the session's last click item's frequencies. Those sessions where the last item distributed in the long-tail part are used to build the datasets. According to the Pareto Principle (the 80/20 rule), the upper bound of the last item frequency is set to 90 and 23 for Yoochoose and Diginetica, respectively. We also set a lower bound of ten for both datasets for more robust evaluation results.

\begin{table*}
\centering
\caption{Performance comparison of different sequential recommendation methods with (w) or without (w/o) Mecos. The best performing methods are \textbf{boldfaced} and the best baseline results are indicated by \underline{underline}.}
\label{tab:comparison}
\resizebox{\textwidth}{!}{
\begin{tabular}{llccccccccccccc}
\toprule
\multicolumn{1}{l}{\multirow{3}{*}{\textbf{Dataset}}} &
  \multicolumn{1}{l}{\multirow{3}{*}{\textbf{Metric}}} &
  \multicolumn{2}{c}{\textbf{GRU4Rec}} &
  \multicolumn{2}{c}{\textbf{NARM}} &
  \multicolumn{2}{c}{\textbf{Caser}} &
  \multicolumn{2}{c}{\textbf{STAMP}} &
  \multicolumn{2}{c}{\textbf{SASRec}} &
  \multicolumn{2}{c}{\textbf{SR-GNN}} & 
  % \multicolumn{1}{c}{\multirow{3}{*}{\textbf{Avg. Improv.}}} \\ 
  \multicolumn{1}{c}{\textbf{Avg.}} \\ 
  \cmidrule(r){3-4} \cmidrule(r){5-6}  \cmidrule(r){7-8} \cmidrule(r){9-10}  \cmidrule(r){11-12} \cmidrule(r){13-14} 
  \multicolumn{1}{c}{}                 & \multicolumn{1}{c}{} & w/o    & w      & w/o    & w      & w/o    & w      & w/o    & w      & w/o    & w      & w/o    & w  & \multicolumn{1}{c}{\textbf{Improv.}}  \\ \midrule %[0.5ex]
\multirow{7}{*}{Steam}      & HR@5        & 0.0860 & 0.2018 & 0.0862 & 0.1850 & 0.0839 & $\textbf{0.2029}$ & 0.0780 & 0.1861 & 0.0621 & 0.1431 & \underline{0.1119} & 0.1878 & 121.33\% \\ 
                                     & HR@10       & 0.1498 & 0.3066 & 0.1578 & 0.3002 & 0.1508 & \textbf{0.3136}  & 0.1392 & 0.2969 & 0.1113 & 0.2358 & \underline{0.1885} & 0.3085 & 98.61\% \\
                                     & HR@20       & 0.2567 & 0.4544 & 0.2785 & \textbf{0.4583} & 0.2634 & 0.4500 & 0.2441 & 0.4539 & 0.2074 & 0.3751 & \underline{0.3137} & 0.4561 & 70.70\% \\
                                     & NDCG@5      & 0.0533 & \textbf{0.1308} & 0.0531 & 0.1206 & 0.0518 & 0.1304 & 0.0486 & 0.1184 & 0.0400 & 0.0955 & \underline{0.0701} & 0.1183 & 129.23\% \\
                                     & NDCG@10     & 0.0737 & 0.1639 & 0.0760 & 0.1561 & 0.0732 & \textbf{0.1651} & 0.0681 & 0.1529 & 0.0557 & 0.1245 & \underline{0.0947} & 0.1650 & 112.60\% \\
                                     & NDCG@20     & 0.1005 & 0.1991 & 0.1062 & 0.1951 & 0.1014 & \textbf{0.1997} & 0.0944 & 0.1920 & 0.0798 & 0.1588 & \underline{0.1261} & 0.1945 & 89.23\% \\
                                     & MRR         & 0.0720 & 0.1412 & 0.0543 & 0.1380 & 0.0728 & \textbf{0.1456} & 0.0689 & 0.1353 & 0.0676 & 0.1138 & \underline{0.0936} & 0.1454 & 95.05\% \\  \midrule
\multirow{7}{*}{Electronic} & HR@5        & 0.0745 & 0.1389 & 0.0843 & 0.1479 & 0.0438 & 0.1300 & 0.0464 & 0.0888 & 0.0394 & 0.1029 & \underline{0.1215} & \textbf{0.1619} & 107.41\% \\
                                     & HR@10       & 0.1357 & 0.2431 & 0.1604 & 0.2546 & 0.0870 & 0.2285 & 0.0924 & 0.1652 & 0.0796 & 0.1858 & \underline{0.1935} & \textbf{0.2591} & 91.10\% \\
                                     & HR@20       & 0.2470 & 0.3970 & 0.2949 & \textbf{0.4158} & 0.1740 & 0.3878 & 0.1794 & 0.2995 & 0.1573 & 0.3217 & \underline{0.3236} & 0.4138 & 70.65\% \\
                                     & NDCG@5      & 0.0457 & 0.0856 & 0.0507 & 0.0931 & 0.0255 & 0.0807 & 0.0286 & 0.0546 & 0.0230 & 0.0618 & \underline{0.0655}
                                     & \textbf{0.0975} & 115.98\% \\
                                     & NDCG@10     & 0.0652 & 0.1180 & 0.0751 & 0.1277 & 0.0393 & 0.0975 & 0.0432 & 0.0817 & 0.0359 & 0.0890 & \underline{0.0950} & \textbf{0.1324} & 95.92\% \\
                                     & NDCG@20     & 0.0930 & 0.1565 & 0.1088 & \textbf{0.1682} & 0.0610 & 0.1522 & 0.0650 & 0.1122 & 0.0553 & 0.1231 & \underline{0.1202} & 0.1678 & 84.53\% \\
                                     & MRR         & 0.0684 & 0.1061 & 0.0699 & 0.1131 & 0.0440 & 0.0903 & 0.0515 & 0.0736 & 0.0514 & 0.0915 & \underline{0.0921} & \textbf{0.1243} & 63.01\% \\  \midrule
\multirow{7}{*}{Tmall}      & HR@5        & 0.2863 & 0.3947 & 0.3105 & 0.4005 & 0.1006 & 0.3384 & 0.2791 & 0.3719 & 0.0862 & 0.3215 & \underline{0.3263} & \textbf{0.4030} & 105.49\% \\
                                     & HR@10       & 0.3437 & 0.4666 & 0.3860 & \textbf{0.4727} & 0.1724 & 0.4023 & 0.3373 & 0.4386 & 0.1307 & 0.3694 & \underline{0.4081}
                                     & 0.4624 & 69.59\% \\
                                     & HR@20       & 0.4371 & 0.5632 & 0.4832 & \textbf{0.5703} & 0.2966 & 0.5070 & 0.4252 & 0.5348 & 0.2086 & 0.4487 & \underline{0.5133} & 0.5616 & 44.68\% \\
                                     & NDCG@5      & 0.2470 & 0.3409 & \underline{0.2747} & \textbf{0.3450} & 0.0640 & 0.2914 & 0.2409 & 0.3245 & 0.0578 & 0.2883 & 0.2548 & 0.3375 & 147.48\% \\
                                     & NDCG@10     & 0.2655 & 0.3641 & \underline{0.2957} & \textbf{0.3680} & 0.0870 & 0.3107 & 0.2597 & 0.3456 & 0.0721 & 0.3040 & 0.2921 & 0.3609 & 116.16\% \\
                                     & NDCG@20     & 0.2889 & 0.3883 & 0.3201 &\textbf{0.3927} & 0.1181 & 0.3339 & 0.2817 & 0.3697 & 0.0915 & 0.3232 & \underline{0.3264} & 0.3848 & 90.36\%\\
                                     & MRR         & 0.2580 & 0.3477 & 0.2848 & 0.3510 & 0.0799 & 0.2996 & 0.2520 & 0.3327 & 0.0659 & 0.2984 & \underline{0.2956} & \textbf{0.3635} & 123.46\% \\\bottomrule

\end{tabular}}
% \begin{tablenotes}
%     \item[1] The best performing method is boldfaced. The underlined number is the second best performing method.
% \end{tablenotes}

\end{table*}

\textbf{Metrics. }
To evaluate the performance of the proposed method, we adopt three common metrics in our experiments, HR@$p$ (Hit Ratio), NDCG@$p$ (Normalized Discounted Cumulative Gain), and MRR (Mean Reciprocal Rank). HR@$p$ is the proportion of cases having the desired item amongst the top-$p$ items in all test cases. It is equivalent to Recall@$p$ and proportional to Precision@$p$. NDCG@$p$ takes the position of correctly recommended items into consideration in the top-$p$ list, where hits at higher positions get higher scores. MRR is the average of reciprocal ranks of the desired items. It is equivalent to Mean Average Precision (MAP).

% \begin{itemize}
% \item {\textbf{HR@$p$}: HR@$p$ (Hit Ratio) is a common metric to evaluate the performance of sequential recommendation models. HR@$p$ is the proportion of cases having the desired item amongst the top-$p$ items in all test cases. It is equivalent to Recall@$p$ and proportional to Precision@$p$.}
% % \item {\textbf{MRR}: MRR (Mean Reciprocal Rank) is the average of reciprocal ranks of the desired items. MRR is especially important to measure the performance of sequential recommendation because it considers the order of recommendation results and users tend to focus on higher-ranked items. And it is equivalent to Mean Average Precision (MAP).}
% \item {\textbf{NDCG@$p$}: NDCG@$p$ (Normalized Discounted Cumulative Gain) takes the position of correctly recommended items into consideration in the top-$p$ list, where hits at higher positions get higher scores.}

% \end{itemize}

% Because most users are only interested in viewing recommendations on the first page of real application scenarios (e.g., web sites of e-commerce), the relevant item should be amongst the first few items in the recommendation list \cite{hu2017diversifying, quadrana2017personalizing}. So we report values of HR and NDCG at $p = \{5, 10, 20\}$.

\textbf{Baselines. }
Because there is no side information in the setting of sequential recommendation task, the previous cold-start methods cannot be applied here (e.g., content-based methods, cross-domain recommenders, and evidence candidate selection strategies). Thus, we focus on models that based only on user-item interactions. We compare proposed framework with following representative and state-of-the-art methods: (1) RNN-based methods including GRU4Rec \cite{hidasi2015session} and NARM \cite{li2017neural}. (2) CNN-based method Caser \cite{tang2018personalized}. (3) Attention-based methods including STAMP \cite{liu2018stamp} and SASRec \cite{kang2018self}. (4) Graph-based method SR-GNN \cite{wu2019srgnn}.

\textbf{Reproducibility Settings. }
We follow the original settings suggested by authors to train all baseline models and the pre-trained embeddings, which makes sure that no side information has been included. Each ground-truth next-click item is paired with 127 negative items ($N=128$) randomly sampled from $\mathcal{Y}_\text{test}$. All models are trained with the same data (pre-train data, all sequences in $\mathcal{T}_\text{meta-train}$ and support sets in $\mathcal{T}_\text{meta-valid}$/$\mathcal{T}_\text{meta-test}$) for fair comparisons. Moreover, we vary the $K$ to investigate the framework performance with different support set sizes and report $K=3$ in default. Besides, the matching step $t$ and hidden dimensionality $d$ is set to 2 and 100, respectively. All hyper-parameters are chosen based on model's performances on $\mathcal{T}_\text{meta-valid}$. We run all models five times with different random seeds and report the average.
\subsection{Results and Analysis}
% \subsection{Comparison of Baseline Methods with and without Mecos (RQ1)}

\textbf{Overall Results (RQ1). }
The experimental results with all baseline methods with or without Mecos are illustrated in Table \ref{tab:comparison}. Mecos significantly improves the state-of-the-art model performance with all metrics in all datasets. That demonstrates the effectiveness of our proposed framework. In addition, since meta-sequences account for a considerable proportion in the whole datasets, Mecos can also help the sequential recommender to perform better in a normal situation (not only in cold-start problem). Besides, Mecos has a larger improvement over original models when $p$ in top-$p$ is smaller. A small value of $p$ in the top-$p$ recommendation means the correct results are in the first few items on the list. As people tend to pay more attention to the items at the top of the page, our framework can help original models to produce more accurate and user-friendly recommendations.

Moreover, with the help of Mecos, the best models in the three datasets are different. In Steam, \emph{Caser with Mecos} achieves the best on five of the seven metrics. Generally, \emph{SR-GNN with Mecos} and \emph{NARM with Mecos} outperform others in Electronic and Tmall, respectively. This indicates that in different recommendation scenarios, we need to adopt different models to get the best effect in alleviating cold-start problems. By incorporating existing base models, Mecos can be utilized as a common solution for different data.

\begin{table}
\centering
\caption{Performance of ablation on different components.}

\scriptsize
\label{tab:ablation}
\resizebox{0.45\textwidth}{!}{
\begin{tabular}{llcccc}
\toprule
\textbf{Dataset}                     & \textbf{Metric} & \textbf{Mecos\_R} & \textbf{Variant\_1} & \textbf{Variant\_2} & \textbf{Variant\_3} \\ \midrule
\multirow{2}{*}{Steam} & HR@10   & \textbf{0.1606} & 0.0933 & 0.1163 & 0.0716 \\
                                
                                % & HR@20   & \textbf{0.2945} & 0.1727 & 0.1973 & 0.1472 \\ [0.5ex]
                                % & NDCG@5  & \textbf{0.0590} & 0.0293 & 0.0387 & 0.0229 \\ [0.5ex]
                                & NDCG@10 & \textbf{0.0805} & 0.0445 & 0.0555 & 0.0339 \\ 
                                % & NDCG@20 & \textbf{0.1136} & 0.0643 & 0.0755 & 0.0390 \\ [0.5ex]
                                & MRR     & \textbf{0.0551} & 0.0405 & 0.0476 & 0.0294 \\ \midrule
\multirow{2}{*}{Electronic} & HR@10   & \textbf{0.1382} & 0.0832 & 0.1207 & 0.0770 \\
                                
                                % & HR@20   & \textbf{0.2531} & 0.1674 & 0.2373 & 0.1582 \\ [0.5ex]
                                % & NDCG@5  & \textbf{0.0463} & 0.0257 & 0.0390 & 0.0214 \\ [0.5ex]
                                & NDCG@10 & \textbf{0.0665} & 0.0382 & 0.0572 & 0.0329 \\ 
                                % & NDCG@20 & \textbf{0.0952} & 0.0592 & 0.0870 & 0.0512 \\ [0.5ex]
                                & MRR     & \textbf{0.0658} & 0.0355 & 0.0505 & 0.0314 \\ \midrule
\multirow{2}{*}{Tmall} & HR@10   & \textbf{0.3481} & 0.3135 & 0.3230 & 0.2921 \\
                                
                                % & HR@20   & \textbf{0.4193} & 0.3760 & 0.3930 & 0.3609 \\ [0.5ex]
                                % & NDCG@5  & \textbf{0.2774} & 0.2505 & 0.2547 & 0.2318 \\ [0.5ex]
                                & NDCG@10 & \textbf{0.2915} & 0.2787 & 0.2737 & 0.2552 \\ 
                                % & NDCG@20 & \textbf{0.3093} & 0.2756 & 0.2812 & 0.2623 \\ [0.5ex]
                                & MRR     & \textbf{0.2891} & 0.2461 & 0.2503 & 0.2324 \\ \bottomrule
\end{tabular}}
\end{table}

% \subsection{Ablation study (RQ2)}
\textbf{Ablation Study (RQ2). }
We conduct an ablation study to investigate the contribution of each component. To remove the influence of pre-trained embedding models, we conduct our framework based on randomly initialized item embeddings (Mecos\_R). And the following variants of it are tested on all datasets, where the results are reported in Table \ref{tab:ablation}:
\begin{itemize}
    \item {\textbf{(Variant\_1)} Mecos\_R without pair encoder. We replace the sequence pair encoder by a simple mean-pooling layer over all item embeddings in a sequence. As shown in Table \ref{tab:ablation}, it performs much worse than Mecos\_R, indicating the importance of pair encoder. }
    
    % Simply using mean pooling may include some noise into users' representation. For example, if users falsely click some disinterested items (called noise click), mean pooling will treat them equal to those other items. Hence, the user's embedding could be less accurate, especially when the user is interested in cold-start items. Our attention-based pooling strategy can determine those noisy clicks, thus generate more representative user's embedding. Without accurate sequence pair representations, our model cannot well utilize the limited training instances in the support set.}
    \item {\textbf{(Variant\_2)} Mecos\_R without matching processor. We set the matching steps to zero to eliminate its effects. From the results, Mecos\_R largely outperforms Variant\_2. That demonstrates the matching-processor has strong ability in computing relevance between query and support set, which might contribute to its ability to extract real-world connections.}
    \item {\textbf{(Variant\_3)} Mecos\_R without pair encoder and matching processor. Under that setting, Variant\_3 generates the recommendation score for each candidate only based on the similarity between the embedding of the query and the support set, which are both generated by a simple mean pooling layer. As might be expected, it has the worst performance of all variants, which further verifies the importance of these key components. However, it is still better than most baseline models, demonstrating the effect of the pure framework without backbone models.}
\end{itemize}

\begin{figure}
\centering
\centering
{\includegraphics[width=\linewidth]{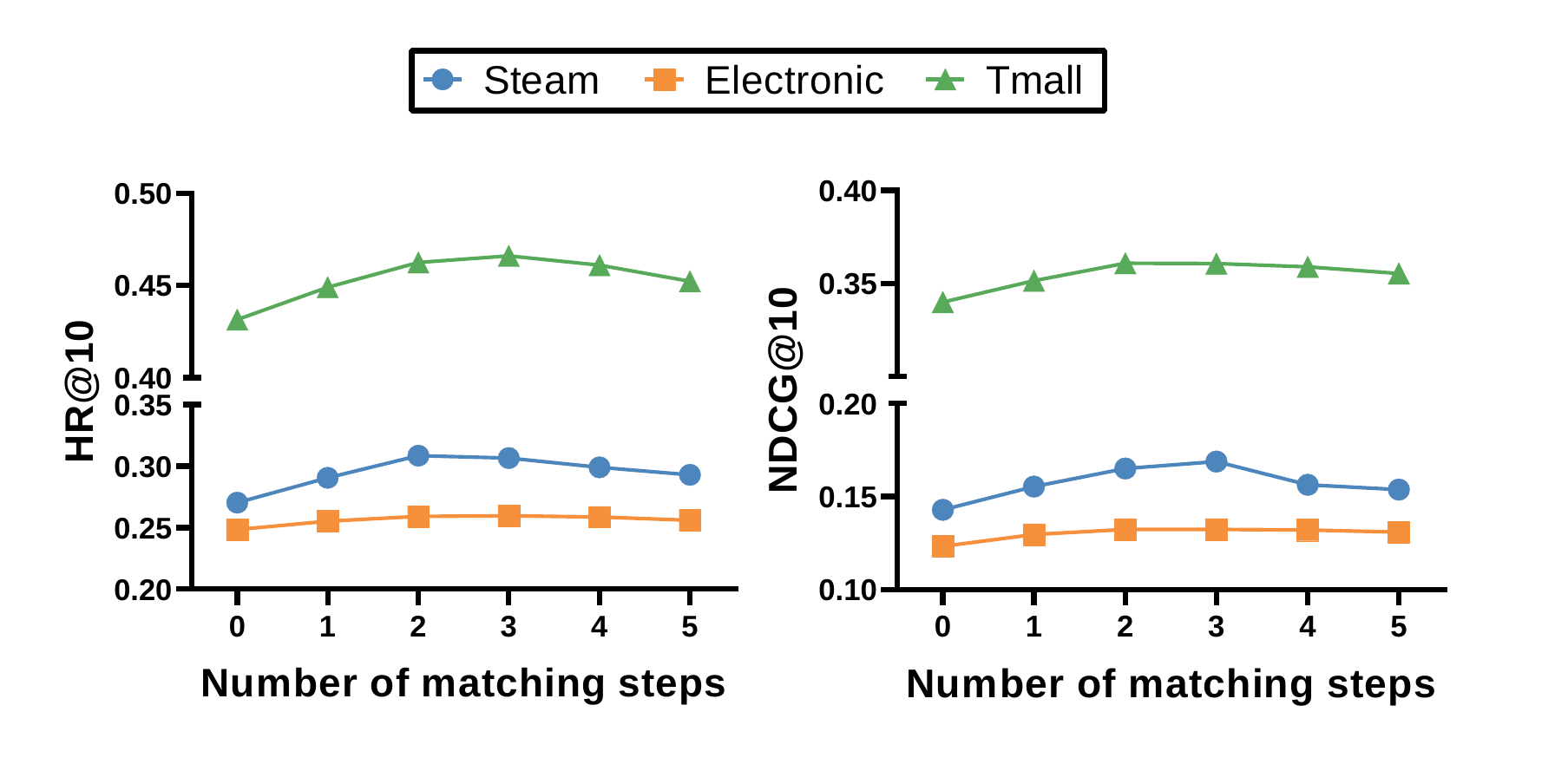}}

%\vspace{-4mm}
\caption{Performance w.r.t. the number of matching steps $t$.}
\label{fig: step}
%\vspace{-2mm}
\end{figure}

% \begin{figure}[htb]
% \centering
% \subfigure[HR@p in Steam]{\includegraphics[width=0.48\linewidth]{step/sh.pdf}}
% \subfigure[NDCG@p in Steam]{\includegraphics[width=0.48\linewidth]{step/sn.pdf}}
% \subfigure[HR@p in Electronic]{\includegraphics[width=0.48\linewidth]{step/eh.pdf}}
% \subfigure[NDCG@p in Electronic]{\includegraphics[width=0.48\linewidth]{step/en.pdf}}
% \subfigure[HR@p in Tmall]{\includegraphics[width=0.48\linewidth]{step/th.pdf}}
% \subfigure[NDCG@p in Tmall]{\includegraphics[width=0.48\linewidth]{step/tn.pdf}}
% %\vspace{-4mm}
% \caption{Performance w.r.t. the number of matching steps $t$.}
% \label{fig: step}
% %\vspace{-2mm}
% \end{figure}

\subsection{Study of Mecos (RQ3)}
% In this section, we conduct experiments with different hyper-parameter settings to study their influence on the framework. We start by exploring the impact of matching steps in the matching network. Moreover, we analyze the effect of hidden dimensionality.

\begin{figure}[tbp]
\centering
\subfigure[SR-GNN]{\includegraphics[width=0.48\linewidth]{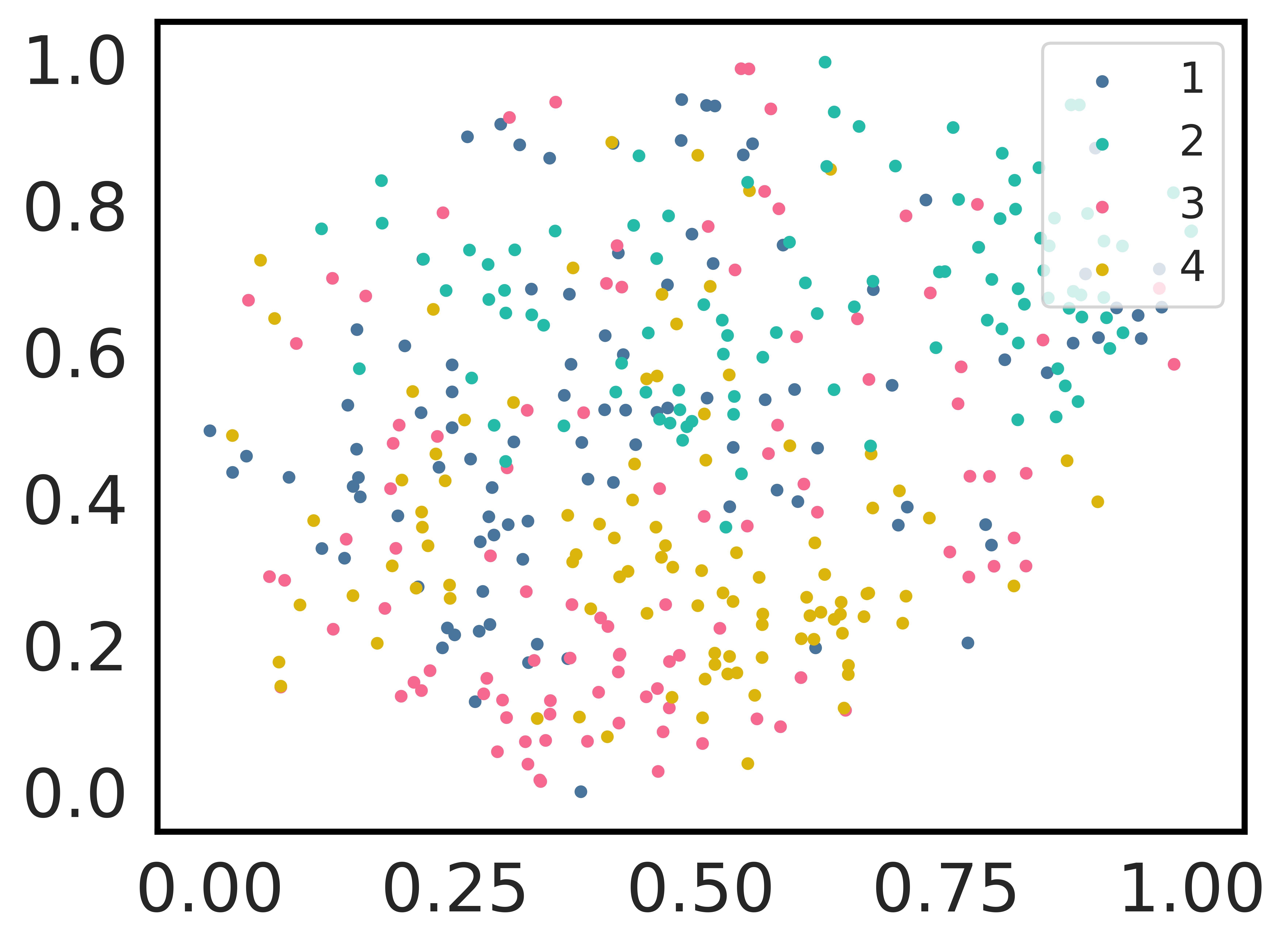}}
\subfigure[SR-GNN with Mecos]{\includegraphics[width=0.48\linewidth]{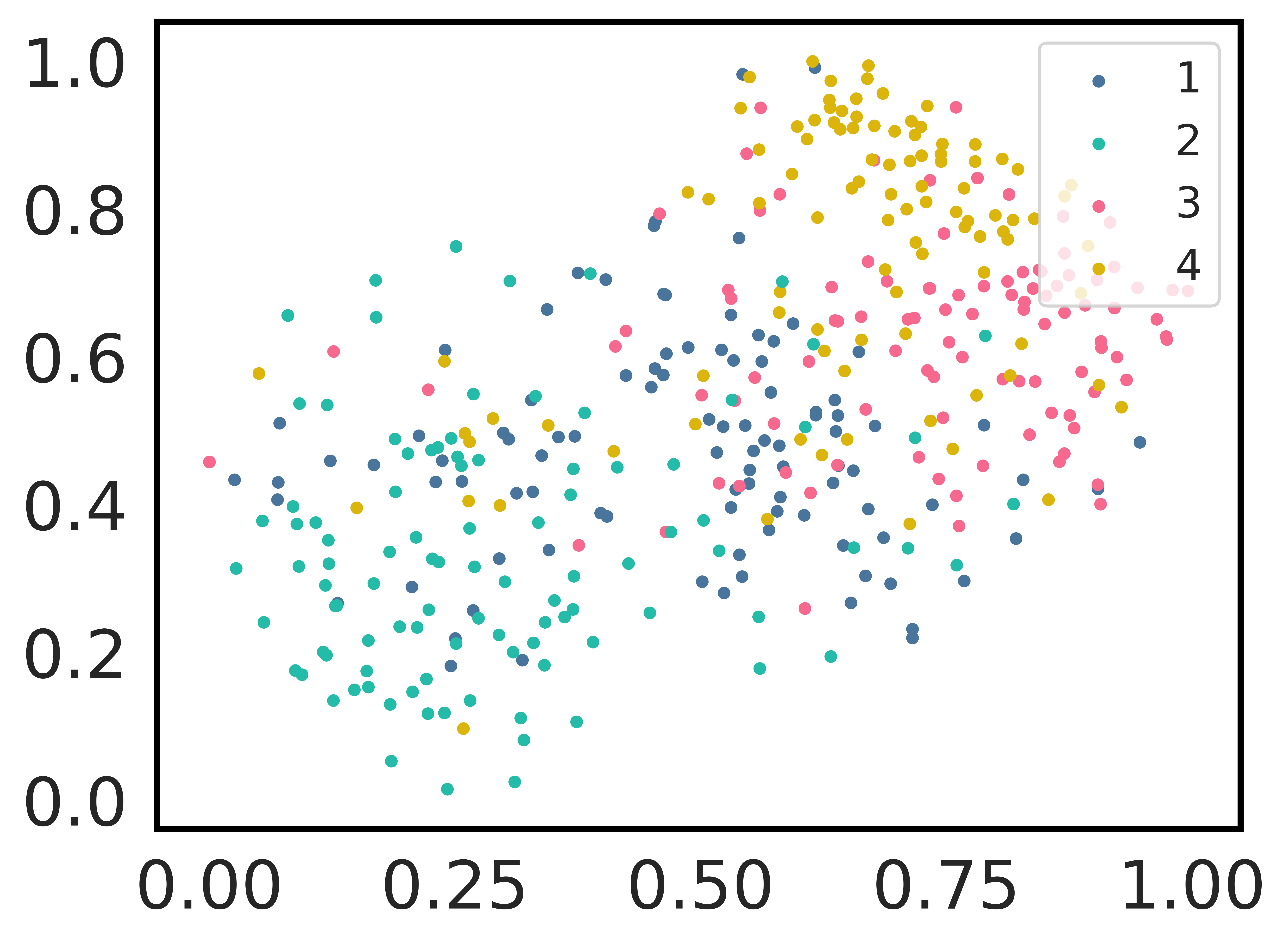}}
% \subfigure[SR-GNN]{\includegraphics[width=0.48\linewidth]{t-sne/avgsrgnn.png}}
% \subfigure[SR-GNN with Mecos]{\includegraphics[width=0.48\linewidth]{t-sne/match2srgnn.png}}
%\vspace{-4mm}
\caption{Embedding visulization of query sequence pairs belonging to different candidate next-click items. Points with the same color denote sequence pairs belonging to the same target item.}
% \caption{Embedding visulization of query sequence pairs $\mathbf{h}_\mathrm{q}$ belonging to different candidate next-click items. Points with the same color denote sequence pairs belonging to the same target item.}
\label{fig: t-sne}
%\vspace{-2mm}
\end{figure}

% \subsubsection {Impact of matching steps}
\textbf{Impact of matching steps. }
As the matching processor contributes a lot to Mecos, it is necessary to further study it. We consider changing the matching steps $t$ to study the influence of matching between query and reference. From Figure \ref{fig: step}, increasing the number of matching steps significantly improves the model performance, which is contributed to the effective similarity matching of the recurrent matching processor. However, it is noteworthy that the performance of the model starts to decrease after we continue to increase matching steps. In general, the performance reaches its best when matching steps are maintained on two or three. The reason for that might be the overfitting due to too many steps. As the matching processor continuously injects information from the support set into the query representation in each matching step, the embedding space can become too crowded, resulting in the hubness problem. We can observe the clustering in the embedding visualization section (Figure \ref{fig: t-sne}).

\begin{figure}[ht]
\centering
{\includegraphics[width=\linewidth]{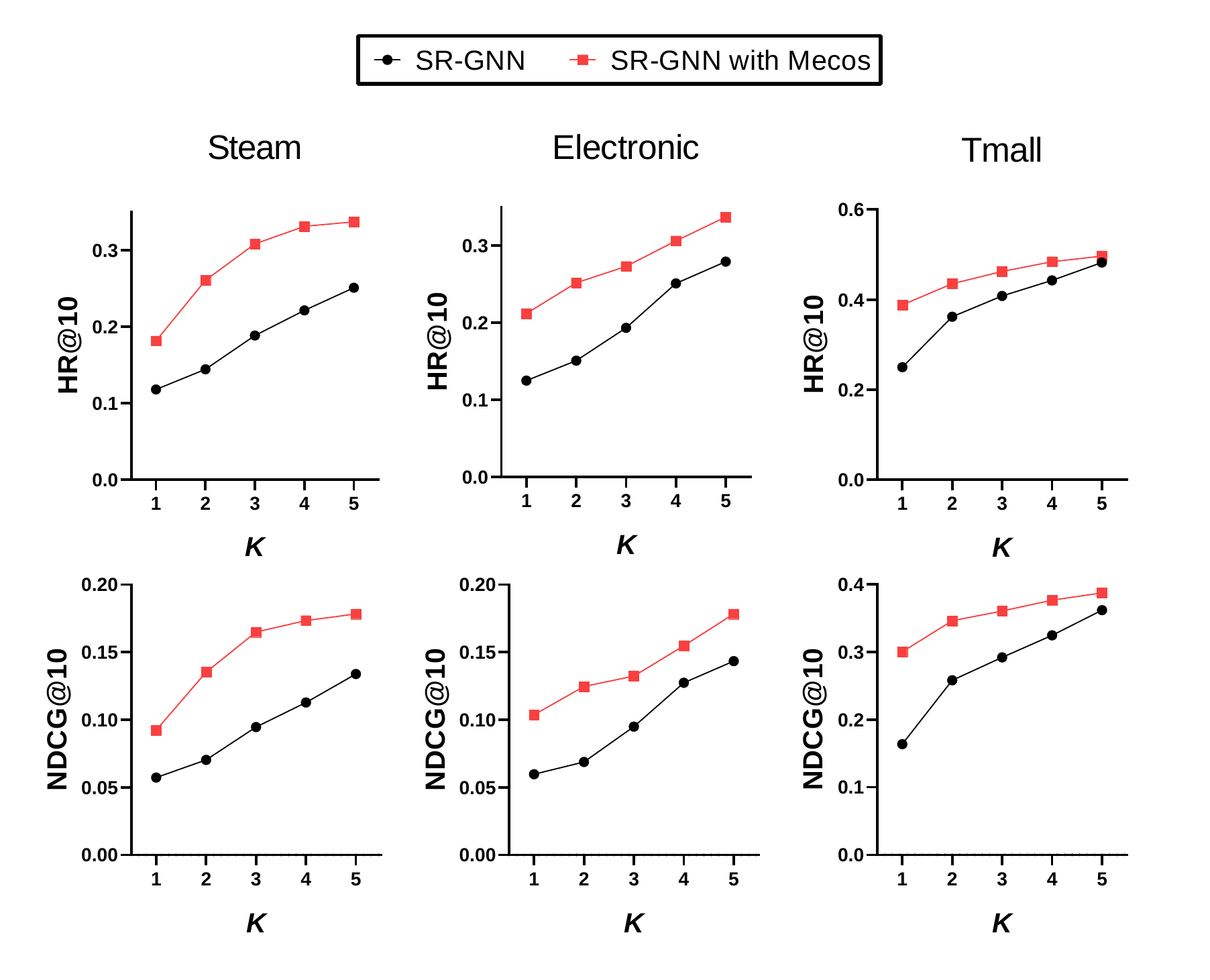}}
%\vspace{-4mm}
\caption{Performance w.r.t. the value of few-shot size $K$. The performance of \emph{SR-GNN with Mecos} consistently outperforms SR-GNN.}
\label{fig: k}
%\vspace{-2mm}
\end{figure}

Besides, by jointly comparing Table \ref{tab:comparison} and Figure \ref{fig: k}, we observe that \emph{SR-GNN with Mecos} is consistently superior to SR-GNN. Even without any matching steps, Mecos can also help the base model in alleviating the item cold-start problem. That again verifies the effectiveness and robustness of the framework.

\textbf{Impact of support set size. }
Support set size $K$ represents how many instances that the model can use for training. To study the impact of it, we test Mecos with the strongest base model, SR-GNN, with different settings of $K$. According to Figure \ref{fig: k}, performance increases with the increase of $K$. That is reasonable because a larger support set contains more known interactions for the target cold-start items. In this way, Mecos can provide a more precise characterization of the product's potential users. And it is noteworthy that Mecos can also help a lot in cold-start item recommendation with only one-shot training instance ($K=1$), illustrating that Mecos produces reliable results even when training instances are extremely limited (e.g., one-shot learning). 

However, for \emph{SR-GNN with Mecos}, the growth trend of three metrics flattens out with the increase of the support set size $K$. And the performance gap between SR-GNN with and without Mecos is narrowing, indicating that Mecos is not suitable for items that have rich known interactions for training. That phenomenon is most obvious in Tmall datasets for it has the lowest proportion of cold-start items.

\subsection{Embedding Visualization (RQ4)}
% In this section, we want to explore how Mecos helps with the representation. And we are also curious about what attributes to the significant improvement of Mecos over the base methods. Thus, we study it in a 2-D embedding space to intuitively study the impact of our framework.

To intuitively study the impact of our framework, we visualize the query sequence pair embedding for different candidate next-click items by t-SNE \cite{tsne}. For each of those items, all others can be seen as negative data points. Thus, a reliable model should be able to distinguish these embeddings after training. We conduct the visualization on Steam with four different target items, each target item has 100 query sequence pairs. And we choose SR-GNN as the base model because it outperforms others in Steam. From Figure \ref{fig: t-sne}, it is clear that the model with Mecos better distinguishes different types of embeddings.
% For base models, we use a simple mean-pooling layer over all item embeddings, pre-trained by base models, in a sequence to get $\mathbf{h}_\mathrm{q}$.

% , which qualitatively verifies that the proposed framework is capable of making accurate recommendations. And sequence pairs belonging to the same candidate next-click item tend to be closer when the model is integrated with Mecos. It indicates that Mecos can improve the model's clustering ability, which benefits the representation a lot.

\section{Conclusion}

This work introduces Mecos to alleviate the item cold-start problem in sequential recommendation, which is an important problem in a novel and challenging context. With the pre-trained item embeddings as the input, our framework could be painlessly integrated with other state-of-the-art models. Mecos performs jointly optimization of the sequence pair encoder and the multi-step matching network. And a meta-learning based gradient descent approach is employed for parameter optimization. Once trained, our framework can be directly adapted to the new items updated in the online environment. Extensive experiments conducted on three real-world datasets illustrate the effect of Mecos and the contribution of each component.

\section{Acknowledgements}
This work is supported by National Key Research and Development Program (2018YFB1402600) and National Natural Science Foundation of China (61772528).

% This work represents an initial step to alleviate item cold-start in sequential recommendation, and our work performs a general framework with high future development potential. For example, by replacing aggregation function in the support set encoding witrcite{wang2019CSRM} or GNN\cite{kipf2016semi}, more valuable information could be extracted from the pre-trained embedding. By utilizing GNN to aggregate few-shot instances in the support set, the hidden structural relation could also be injected in the representation of support set effectively. And taking advantage of a different meta-training process could also improve learning and optimization. Moreover, our framework can also integrate other forms of structural information, such as knowledge graph, to increase the quality of embeddings and extend to other recommendation scenarios.

\bibliography{sample-base}

\end{document}